\documentclass[aps,prb,reprint,superscriptaddress,showpacs]{revtex4-1}

\usepackage{graphicx}
\usepackage{dcolumn}
\usepackage{bm}

\begin{document}


\title{Spin current and second harmonic generation in non-collinear magnetic systems: the hydrodynamic model}


\author{E. A. Karashtin}
\email{eugenk@ipmras.ru}
\author{A. A. Fraerman}
\affiliation{Institute for Physics of Microstructures RAS, GSP-105, 603950, Nizhny Novgorod, Russia}
\affiliation{University of Nizhny Novgorod, 23 Prospekt Gagarina, 603950, Nizhny Novgorod, Russia}


\date{\today}

\begin{abstract}
We report a theoretical study of the second harmonic generation in a noncollinearly magnetized conductive medium with equilibrium spin current. The hydrodynamic model is used to unravel the mechanism of a novel effect of the double frequency signal generation due to the spin current. According to our calculations, this second harmonic response appears due to the ``non-adiabatic'' spin polarization of the conduction electrons induced by the oscillations in the non-uniform magnetization forced by the electric field of the electromagnetic wave. Together with the linear velocity response this leads to the generation of the double frequency spin current. This spin current is converted to the electric current via the inverse Spin Hall effect, and the double-frequency electric current emits the second harmonic radiation. Possible experiment for detection of the new second harmonic effect is proposed.
\end{abstract}


\maketitle


\section{Introduction\label{Intro}}
Spin current is the flow of spin. It is often accompanied by the flow of charge, i.e. the electric current \cite{2Zutic}. Systems with such current are widely studied due to the possibility to electrically switch the magnetic state \cite{5Stamps, 6Katine, 7Parkin, 8Atkinson}. This mechanism is commonly used when the spin injection is needed \cite{200Sato, 36Johnson, 17Shikin}. Another way to produce the spin current is spin pumping \cite{9Tserkovnyak, 10Brataas}. Here, the spin current is created with no average charge motion. Such spin flow is called the \textit{pure} spin current \cite{12Sonin}. An important effect that appears due to the pure spin current is the inverse spin Hall effect \cite{20Sinova, 25Saitoh, 30Valenzuela, 33Zhao}. It consists in the appearance of the charge current which is perpendicular to the externally created spin current. The mechanism of the inverse spin Hall effect is the spin-orbit interaction \cite{20Sinova}. It provides the posasibility to detect the pure spin current. Other possibilities include the x-ray magnetic circular dichroism measurements \cite{34Li}, the electron spin resonance method \cite{35Liang}, or the methods that utilize the spin-valve configuration \cite{36Johnson, 37Lou, 38Tombros, 39Erve}.

A special kind of systems with the spin current is a system where the pure spin current exists in equilibrium \cite{12Sonin, 49Wang, 41Konig, 45Schutz, 46Heurich, 47Splettstoesser, 48Wang}. One may show that if the magnetic system is considered the magnetization distribution should be non-collinear in order to have an equilibrium spin current \cite{49Wang, 40Chshiev, 42Chen, 43Shen, 44Wang}. Such systems attract attention due to new interesting properties that may be caused by the spin current. Indeed, both the inversion symmetry and the time reversal symmetry are broken in equilibrium here. This leads to several new phenomena. The flexo-magnetoelectric effect was predicted in non-collinearly magnetized media \cite{60Baryakhtar, 61Katsura, 62Bruno, 63Mills, 64Logginov, 65Veshchunov}. It will hopefully provide a new way to control the magnetization without applying a magnetic field. Another predicted effect is the second-order nonlinear optical effect due to the spin current \cite{99Wang}. It was observed in the GaAs semiconductor where the spin current was induced by simultaneous illumination with two laser pulses \cite{100Werake}. Such effect may be used as a powerful tool for spin current diagnostics. This paper is devoted to the investigation of the described second harmonic generation effect in a non-uniformly magnetized medium with equilibrium pure spin current. The microscopic reasons of this effect are discussed in the framework of the simple hydrodynamic model describing the conductance electrons. Possible experiment for detection of the predicted effect is proposed based on the polarization properties of the second harmonic signal.

\section{Symmetry considerations}

We consider a magnetic medium with the magnetization $\mathbf{M}\left(\mathbf{r}\right)$ normalized to unity ($\left|\mathbf{M}\right| = 1$). We find the exchange equilibrium spin current $\sigma_{jk}$ . Such current should transform under a coherent rotation of $\mathbf{M}$ so that its spin component is rotated with $\mathbf{M}$. In this approximation, the  tensor $\sigma_{jk}$ may be written as \cite{300Landau, 62Bruno}:
\begin{equation} \label{Eq1_1}
\sigma_{jk} = u \left[\mathbf{M} \times \frac{\partial \mathbf{M}}{\partial x_k}\right]_j,
\end{equation}
where $u$ is a constant; $j$ is the spin index of the spin current tensor, $k$ is its coordinate index.  Here we suppose that the magnetization changes slowly in space ($\left|\frac{\partial \mathbf{M}}{\partial x_k}\right| << \frac{1}{a}$, $a$ being the lattice constant in the medium, $x_k$ the k-th coordinate in the Cartesian coordinate system). So we take into account only the first-order derivative of $\mathbf{M}$ and neglect the higher derivatives. The polarization at double frequency may be obtained in the following form:
\begin{equation} \label{Eq1_2}
P^{2 \omega}_i = \beta_{ijklp} \sigma_{jk} E^{\omega}_l E^{\omega}_p,
\end{equation} 
here  ${E}^{\omega}_l$ (${E}^{\omega}_p$) denotes the l-th (p-th) component of the electric field $\mathbf{E}^{\omega}$ of the wave propagating in the medium, $\beta_{ijklp}$ is the fifth rank tensor. In an isotropic medium the form of $\beta_{ijklp}$ is restricted to:
\begin{eqnarray} \label{Eq1_3}
\beta_{ijklp} = &A_1& e_{ijk} \delta_{lp} + A_2 e_{ljk} \delta_{ip} \\ &+& A_3 e_{ijl} \delta_{kp} + A_4 e_{ikl} \delta_{jp}, \nonumber
\end{eqnarray}
assuming that the tensor should be symmetric with respect to the $l$ and $p$ indices (the corresponding symmetrizing terms are omitted). The $e_{ijk}$ tensor in (\ref{Eq1_3}) is the completely antisymmetric Levi-Civita tensor, $\delta_{lp}$ is the Kronecker delta. Compiling the formula (\ref{Eq1_3}) with (\ref{Eq1_1}) and (\ref{Eq1_2}) we have for this case:
\begin{eqnarray} \label{Eq1_4}
\mathbf{P}^{2 \omega} = &B_1& \mathbf{P} \left(\mathbf{E}^{\omega}\right)^2 + B_2 \mathbf{E}^{\omega} \left(\mathbf{P} \cdot \mathbf{E}^{\omega}\right) \\ &+& B_3 \left[\mathbf{E}^{\omega} \times \left[\mathbf{M} \times \left(\mathbf{E}^{\omega} \cdot \nabla\right) \mathbf{M}\right]\right],\nonumber
\end{eqnarray}
where $B_l$ are the constants corresponding to $A_l$ ($l=1..3$), the term with $A_4$ is linearly dependent on that with $A_1 ..A_3$ and therefore is omitted, $\mathbf{P}$ denotes the electric polarization that appears in a non-uniformly magnetized medium due to the flexo-magnetoelectric effect \cite{61Katsura}:
\begin{equation} \label{Eq1_5}
\mathbf{P} = \alpha e_{ijk} \left[\mathbf{M} \times \frac{\partial \mathbf{M}}{\partial x_j}\right]_{i} \mathbf{e}_k,
\end{equation}
$\alpha$ is a constant, $\mathbf{e}_k$ stands for the k-th unit vector in the Cartesian coordinate system.

It should be noted that the formula (\ref{Eq1_2}) is not invariant with respect to the rotation of $\mathbf{M}$. Therefore the second-harmonic generation due to the spin current may be obtained only if the spin-orbit interaction is taken into account in addition to the exchange interaction \cite{201Landau}.

\section{Model and Equations}

In order to understand the origin of the effect described by (\ref{Eq1_2}) we consider the motion of the conduction electrons with the hydrodynamic equations \cite{202Boyd} extended with the equations for spin \cite{20Sinova, 110Dyakonov, 111Dyakonov, 112Shpiro}.

The Euler equation for the total electron velocity $\mathbf{V}$ reads
\begin{eqnarray} \label{Eq2_1}
 \frac{\partial \mathbf{V}}{\partial t} &+& \left(\mathbf{V} \cdot \nabla\right) \mathbf{V} = -\frac{e}{m} \mathbf{E}^{\omega} - \frac{e}{m} \mathbf{E} \\ &+& \eta_H \left[\mathbf{V} \times \mathbf{M}\right] -\frac{e}{mc}\left[\mathbf{V} \times \mathbf{B}^{\omega}\right] - \frac{\mathbf{V}}{\tau_p}, \nonumber
\end{eqnarray}
where $\mathbf{B}^{\omega}$ is the magnetic field of the wave, $\mathbf{E}$ is the field induced by the wave, $e$ is the absolute electron charge, $m$ is its mass, $\eta_H$ is the anomalous Hall constant (that has its roots in the spin-orbit interaction), $\tau_p$ is the momentum relaxation time, $c$ is the light velocity. The magnetic moment induced by the conduction electrons is neglected in (\ref{Eq2_1}). Besides, we do not take into account the spatial dispersion, and thus have $\mathbf{E}^{\omega} = \mathbf{E}_0 \exp \left(i \omega t \right)$, $\omega$ is the wave frequency. (The Lorentz force is written in (\ref{Eq2_1}) with $\mathbf{B}^{\omega}$ representing an additional nonlinearity mechanism; however, this mechanism is not important in the scope of current paper.)

Equation (\ref{Eq2_1}) is accompanied by the standard law of conservation of mass:
\begin{equation} \label{Eq2_2}
\frac{\partial n}{\partial t} + div \mathbf{q} = 0,
\end{equation}
where $n$ is the electron density, and the electron flow $\mathbf{q}$ is defined as:
\begin{equation} \label{Eq2_3}
\mathbf{q} = n \mathbf{V} = n \mathbf{v} + \alpha_{SH} e_{ijk} \sigma_{ij} \mathbf{e}_k.
\end{equation}
Here, a phenomenological term is added that describes the inverse spin Hall effect (which appears due to the spin-orbit interaction), $\alpha_{SH}$ is the inverse spin Hall effect constant. In (\ref{Eq2_3}),  $\mathbf{v}$ is a "normal" electron velocity in the absence of the inverse spin hall effect (added for the convenience), $e_{ijk}$ is the completely antisymmetric rank three tensor. It is important to note that we neglect the diffusion term in (\ref{Eq2_3}). One may check that the diffusion terms lead to higher-order derivatives of $\mathbf{M}$ in an isotropic medium with no boundaries.

The induced electric field $\mathbf{E}$ in (\ref{Eq2_1}) is determined by the Maxwell equation:
\begin{equation} \label{Eq2_4}
div \mathbf{E} = -4 \pi e \left(n - n_0\right),
\end{equation}
where $n_0$ is the equilibrium electron density (the wave field $\mathbf{E}^{\omega}$ is neglected since we do no take into account the spatial dispersion). The spin current tensor $\sigma_{ij}$ is determined as follows:
\begin{equation} \label{Eq2_5}
\sigma_{ij} = s_i v_j + A \left[\mathbf{M} \times \frac{\partial \mathbf{M}}{\partial x_j} \right]_i	,
\end{equation}
where we phenomenologically add the equilibrium exchange spin current that is determined by the constant $A$; $s_i$ in (\ref{Eq2_5}) stands for the i-th component of the electron spin density $\mathbf{s}$. We are seeking the effect linear with respect to the spin-orbit interaction. Hence there is no spin Hall effect in the equation for spin current (\ref{Eq2_5}); the diffusion term is also neglected in (\ref{Eq2_5}).
In order to make the system of equations that describe our medium full we add the spin conservation law \cite{12Sonin}:
\begin{equation} \label{Eq2_6}
\frac{\partial \mathbf{s}}{\partial t} + \frac{\partial \sigma_{ij}}{\partial x_j} \mathbf{e_i} + \frac{\left[ \mathbf{s} \times \mathbf{M} \right]}{\tau_{ex}} + \frac{\mathbf{s} - \mathbf{s}_0}{\tau_s} = 0.
\end{equation}
Here $\tau_{ex} = \frac{\hbar}{2 J}$ is the exchange time that determines the spin precession period ($J$ is the exchange constant), $\tau_s$ is the spin relaxation time, and $\mathbf{s}_0 = \beta \mathbf{M}$ is the equilibrium average spin density of the electrons ($\beta \approx n_0 \frac{J}{\varepsilon_F}$ is a constant that characterizes the difference in the density of spin-up and spin-down electrons in the ferromagnet, $\varepsilon_F$ is the Fermi energy).

We solve the system of equations (\ref{Eq2_1})~--~(\ref{Eq2_6}) (see Appendix). The second harmonic electric polarization is obviously connected to the double frequency electron flow $\mathbf{q}^{2 \omega}$:
\begin{equation} \label{Eq2_7}
\mathbf{P}^{2 \omega} = -\frac{e}{2 i \omega} \mathbf{q}^{2 \omega}.
\end{equation}
The main approximations and restrictions of our model are summed in the list below.
\renewcommand{\labelenumi}{\theenumi}
\renewcommand{\theenumi}{\roman{enumi}}
\begin{enumerate}
\item \label{it1} The localized electrons induce the non-uniform magnetization $\mathbf{M}\left(\mathbf{r}\right)$. We restrict ourselves to the first order derivatives of $\mathbf{M}\left(\mathbf{r}\right)$, neglecting higher order derivatives.
\item \label{it2} The delocalized conduction electrons are in charge of the optical response, including the non-linear effects.
\item \label{it3} Only the linear in $\alpha_{SH}$ and $\eta_{H}$ (linear in the spin-orbit interaction) terms are taken into account.
\end{enumerate}

\section{Calculations and Discussion}

We find the second harmonic electron flow that is proportional to the second order of $\mathbf{M}$ and hence may be determined by the spin current. It has the following form:
\begin{eqnarray} \label{Eq3_1}
\mathbf{q}^{2 \omega} = w \left[\mathbf{E}^{\omega} \times \left[\mathbf{M} \times \left(\mathbf{E}^{\omega} \cdot \nabla\right) \mathbf{M}\right]\right],
\end{eqnarray}
where the constant $w$ is linear within the spin-orbit interaction (proportional to $\alpha_{SH}$, see below). Obviously, the relation (\ref{Eq3_1}) corresponds to the third term in the phenomenological expression (\ref{Eq1_4}) (determined by $B_3$). The terms determined by the static electric polarization $\mathbf{P}$ do not exist in bulk in our hydrodynamic model. We suppose that similar terms should appear if the boundaries are considered.

Calculations show that the constant $w$ is the following (for more details, see Appendix):
\begin{eqnarray} \label{Eq3_2}
w = \frac{-\left(\frac{e}{m}\right)^2 \tau_{ex} \beta \alpha_{SH}}{\left(1 + \left(i \omega + \frac{1}{\tau_s}\right)^2\tau_{ex}^2\right) \left(\left(\omega - \frac{\omega_p^2}{\omega}\right) - \frac{i}{\tau_p}\right)^2},
\end{eqnarray}
here $\omega_p = \frac{4 \pi n_0 e^2}{m}$ is the electron plasma frequency. As expected, it is proportional to the inverse spin Hall effect constant. The mechanism of its appearance follows from the calculations. The electrons oscillation in the electric field of the electromagnetic wave leads to a ``non-adiabatic'' spin polarization that is proportional to $\left[\mathbf{M} \times \left(\mathbf{E}^{\omega} \cdot \nabla\right) \mathbf{M}\right]$. It appears in addition to the ``adiabatic'' transfer of the spin by the moving (oscillating) electrons due to the precession about the local magnetization that is not parallel to this spin in the non-equlilibrium state. Such ``non-adiabatic'' spin is similar to that described by Aharonov and Stern for the electrons in equilibrium \cite{155Aharonov}. Together with the velocity that is proportional to $\mathbf{E}^{\omega}$ the ``non-adiabatic'' spin polarization leads to the second harmonic spin current. The inverse spin Hall effect converts it to the double frequency electric current. The resonance at the plasma frequency appears due to the oscillation of the charge density in the right side of the equation (\ref{Eq2_4}). Note that although the inversion symmetry is broken due to the equilibrium spin current in the system (which is proved by the symmetry consideration) the mechanism described here is not connected directly to the equilibrium spin current and contrarily appears from the dynamic spin current.

\begin{figure}[t]
\includegraphics[width = 2.5in, keepaspectratio=true]{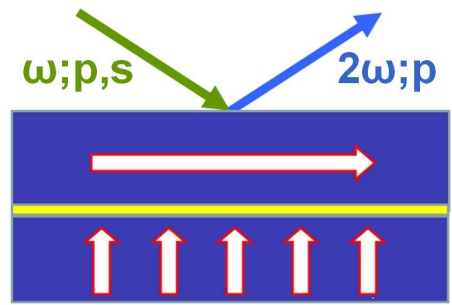}
\caption{\label{Fig_1} (Color online) Possible scheme of the second harmonic generation effect in a non-collinear magnetic structure.}
\end{figure}

The considered effect may be observed in an artificial magnetic system that consists of two ferromagnetic layers with different types of anisotropy divided by a thin non-magnetic interlayer (Figure \ref{Fig_1}). 
Such system has a non-collinear magnetization distribution in equilibrium. Besides, there may be the exchange interaction between two subsystems if the interlayer is thin enough \cite{210Demidov}. The magnetization distribution may be controlled by applying an external magnetic field. The spin current here may be written as $\sigma_{ij} = \tilde{u} \left[\mathbf{M}_1 \times \mathbf{M}_2\right]_j n^{12}_i$ where $\mathbf{n}^{12}$ is a vector directed from the first layer to the second one (along the normal), $\mathbf{M}_{1,2}$ are the magnetizations of the layers, $\tilde{u}$ is a constant. Therefore, we have an $\left[\mathbf{M}_1 \times \mathbf{M}_2\right]$ vector in addition to $\mathbf{M}_1$ and $\mathbf{M}_2$. The polarization properties of the second harmonic generation from a surface for different directions of the magnetization vector are outlined in \cite{220Rzhevsky}. It leads from these properties that if the magnetizations of both layers lie in the pump plane of incidence there is no second harmonic signal in the p-polarization from the first or the second layer separately. However, the new effect that is determined by $\left[\mathbf{M}_1 \times \mathbf{M}_2\right]$ would give a p-polarized second harmonic wave for both s- and p-polarized pump. Thus, the p-p or s-p combination of pump and second harmonic polarization is optimal for observation of the effect of second harmonic generation due to the spin currents in such system.

\section{Conclusion\label{Sum}}

In summary, we theoretically investigate the second harmonic generation effect in a system with non-collinear magnetization distribution. The mechanism of the non-linear optical effect due to the spin current in such system is unraveled. It consists in the ``non-adiabatic'' spin polarization of the conduction electrons which appears due to the oscillations of the electrons in the non-uniform magnetization forced by the electric field of the electromagnetic wave. This spin leads to the second harmonic spin current that is converted to the electric current via the inverse spin Hall effect. The analysis of the polarization properties of the effect shows that it could be observed in a multilayer non-collinear magnetic structure in a p-p or s-p combination of pump and second harmonic polarization.

\begin{acknowledgments}
The authors are thankful to T.~V.~Murzina for valuable discussions. This work was supported by the Russian Science Foundation (Grant No. 16-12-10340).
\end{acknowledgments}

\appendix*
\section{The solution of the equations in the extended hydrodynamic model}

We solve the system of equations (\ref{Eq2_1})~--~(\ref{Eq2_6}). The expansion of the variables that are involved into (\ref{Eq2_1})~--~(\ref{Eq2_6}) as a power series of the electric field of the wave reads:
\begin{eqnarray} \label{Eq2_10}
 n &=& n^0 + n^{\omega} + n^{2 \omega} + ... \,, \\ \label{Eq2_11}
\mathbf{v} &=& \mathbf{v}^0 + \mathbf{v}^{\omega} + \mathbf{v}^{2 \omega} + ... \,, \\ \label{Eq2_12}
\mathbf{s} &=& \mathbf{s}^0 + \mathbf{s}^{\omega} + \mathbf{s}^{2 \omega} + ... \,, \\ \label{Eq2_13}
\mathbf{q} &=& \mathbf{q}^0 + \mathbf{q}^{\omega} + \mathbf{q}^{2 \omega} + ... \,, \\ \label{Eq2_14}
\sigma_{ij} &=& \sigma_{ij}^0 + \sigma_{ij}^{\omega} + \sigma_{ij}^{2 \omega} + ... \,.
\end{eqnarray}
Here we assume that the second-order terms oscillate at the double frequency, i.e. neglect the rectification terms that are beyond the scope of current paper. The terms linear with respect to the small parameters $\alpha_{SH}$ and $\eta_H$ (which are supposed to be of the same order of value due to their spin-orbit roots) are found. We restrict ourselves to the second order with respect to the wave electric (or magnetic) field and to the first order with $\alpha_{SH}$ and $\eta_H$. Besides, we leave only $\mathbf{M}$ and its first-order spatial derivative, neglecting all higher-order derivatives and the powers of the first-order derivative. In these approximations, the solution (\ref{Eq2_10})-(\ref{Eq2_14}) of the equations ({\ref{Eq2_1})-(\ref{Eq2_6}) restricted to the second order in $\mathbf{E}^{\omega}$ are the following.
\begin{widetext}
\begin{eqnarray}
n^0 &=& n_0,\; \mathbf{v}^0 = -\frac{\alpha_{SH} A}{n_0} e_{ijk} \left[\mathbf{M} \times \frac{\partial \mathbf{M}}{\partial x_j}\right]_i \mathbf{e}_k,\; \mathbf{s}^0 = \mathbf{s}_0 = \beta \mathbf{M},\; \sigma_{ij}^0 = A \left[\mathbf{M} \times \frac{\partial \mathbf{M}}{\partial x_j}\right]_i, \; 
\mathbf{q}^0 = 0; \\ \nonumber
\mathbf{v}^{\omega} &=& \frac{-\frac{e}{m}}{i \omega + \frac{1}{\tau_p} + \frac{\omega_p^2}{i \omega}} \left( \mathbf{E}^{\omega} + \frac{\eta_{H} + \frac{\alpha_{SH} \beta}{n_0} \frac{\omega_p^2}{i \omega}}{i \omega + \frac{1}{\tau_p} + \frac{\omega_p^2}{i \omega}} \left[\mathbf{E}^{\omega} \times \mathbf{M}\right]\right), \\
\mathbf{s}^{\omega} &=& \frac{\beta \tau_{ex} \frac{e}{m}}{\left(i \omega + \frac{1}{\tau_p} + \frac{\omega_p^2}{i \omega}\right) \left(1 + \tau_{ex}^2 \left(i \omega + \frac{1}{\tau_s}\right)^2\right)} \left( \tau_{ex}\left(i \omega + \frac{1}{\tau_s}\right) \left(\mathbf{E}^{\omega} \cdot \nabla\right) \mathbf{M} + \left[\mathbf{M} \times \left(\mathbf{E}^{\omega} \cdot \nabla \right) \mathbf{M} \right] \right), \\ \nonumber
n^{\omega} &=& \frac{\frac{e}{m} \left(\mathbf{E^{\omega}} \cdot curl \mathbf{M}\right)}{i \omega \left(\frac{1}{\tau_p} + i \omega\right)} \frac{\alpha_{SH} \beta \left(i \omega + \frac{1}{\tau_p}\right) - n_0 \eta_H}{{i \omega + \frac{1}{\tau_p} + \frac{\omega_p^2}{i \omega}}}, \; \sigma_{ij}^{\omega} = \left(\mathbf{s}_0\right)_i \left(\mathbf{v}^{\omega}\right)_j,\; \mathbf{q}^{\omega} = n_0 \mathbf{v}^{\omega} + \alpha_{SH} \left[\mathbf{s}_0 \times \mathbf{v}^{\omega}\right]; \\ \nonumber
\mathbf{v}^{2 \omega} &=& Q_1 \left[\mathbf{E}^{\omega} \times \left[\left(\mathbf{E}^{\omega} \cdot \nabla\right) \mathbf{M}\right]\right] + Q_2 \left[\mathbf{E}^{\omega} \times \mathbf{B}^{\omega}\right] + Q_3 \left[\left[\mathbf{E}^{\omega} \times \mathbf{B}^{\omega}\right] \times \mathbf{M}\right], \\ \label{EqA_3}
Q_1 &=& \frac{-\left(\frac{e}{m}\right)^2 \left(\eta_H + \frac{\alpha_{SH} \beta}{n_0} \left( \frac{\omega_p^2}{i \omega} - i \omega -\frac{1}{\tau_p} \right)\right) }{\left(i \omega + \frac{1}{\tau_p}+\frac{\omega_p^2}{i \omega}\right)^3 \left(2 i \omega + \frac{1}{\tau_p}+\frac{\omega_p^2}{2 i \omega}\right)}, \;
Q_2 = \frac{\left(\frac{e}{m}\right)^2}{c \left(i \omega + \frac{1}{\tau_p}+\frac{\omega_p^2}{i \omega}\right) \left(2 i \omega + \frac{1}{\tau_p}+\frac{\omega_p^2}{2 i \omega}\right)}, \\ \nonumber
Q_3 &=& \frac{-\left(\frac{e}{m}\right)^2 \left(\eta_H +  \alpha_{SH} \beta \frac{\omega_p^2}{i \omega n_0}\right)}{c \left(i \omega + \frac{1}{\tau_p}+\frac{\omega_p^2}{i \omega}\right)^2 \left(2 i \omega + \frac{1}{\tau_p}+\frac{\omega_p^2}{2 i \omega}\right)} \frac{3 i \omega + \frac{2}{\tau_p}}{2 i \omega + \frac{1}{\tau_p}},\\ \nonumber
\mathbf{s}^{2 \omega} &=& 0, \;
n^{2 \omega} = -\frac{1}{2 i \omega} div \left( n_0 \mathbf{v}^{2 \omega} + \alpha_{SH} \beta \left[\mathbf{M} \times \mathbf{v}^{2 \omega}\right] \right), \;
\sigma_{ij}^{2 \omega} =  \left(\mathbf{s}_0\right)_i \left(\mathbf{v}^{2 \omega}\right)_j +  \left(\mathbf{s}^{\omega}\right)_i \left(\mathbf{v}^{\omega}\right)_j, \\ \nonumber
\mathbf{q}^{2 \omega} &=& n_0 \mathbf{v}^{2 \omega} + n^{\omega} \mathbf{v}^{\omega} + \alpha_{SH} \left( \left[\mathbf{s}_0 \times \mathbf{v}^{2 \omega}\right] + \left[\mathbf{s}^{\omega} \times \mathbf{v}^{\omega}\right] \right).
\end{eqnarray}
\end{widetext}
Although we find the solution in bulk, we should take into acount the boundaries of the sample in order to obtain these results. In this case, the static polarization $\mathbf{P}$ which describes the flexo-magnetoelectric effect and the electric field connected to it via the Maxwell equations compensates the flow that appears due to the inverse spin Hall effect. The ``normal'' part of electron velocity $\mathbf{v}^0$ is therefore nonzero. However, full electron velocity and the electron flow is zero.

It is important to note that the first term in $\mathbf{s}^{\omega}$ linear in $\mathbf{M}$ corresponds to the adiabatic spin transfer under the application of the oscillating electric field to the non-uniformly magnetized medium. The second term in $\mathbf{s}^{\omega}$ that is quadratic with respect to $\mathbf{M}$ appears due to the non-adiabatic processes in the medium.

The expression for $\mathbf{q}^{2 \omega}$ contains the terms linear in $\mathbf{M}$. First, those are the terms that do not contain the differentiation operator and hence may appear in a uniformly magnetized medium. These terms arise from the Lorentz force due to the magnetic field of the wave in an infinite system. Next, there are terms that contain the derivative of $\mathbf{M}$ and appear in a non-uniformly magnetized media. These terms are usually attributed to the toroidal moment of the system \cite{301Petukhov, 150Krutyanskiy, 151Kolmychek}. Besides, the fourth term in its right-hand part ($\alpha_{SH} \left[\mathbf{s}^{\omega} \times \mathbf{v}^{\omega}\right]$) gives the second-order in $\mathbf{M}$ term that is connected to the spin current in the system. The $w_1$ coefficient in (\ref{Eq3_1}) determining it may be easily obtained from the expression for $\mathbf{q}^{2 \omega}$ in (\ref{EqA_3}). It appears due to the non-adiabatic processes mentioned below.

\bibliography{shg}

\end{document}